\documentclass[a4paper,11pt]{article}
\usepackage{graphicx}
\usepackage[english]{babel}
\usepackage{amssymb}
\usepackage{amsmath}

\begin{document}
\title{Phase diagram of semi-hard-core bosons on a square lattice}

\author{V.V. Konev \and Yu.D. Panov}
\date{}

\maketitle

\vspace{-2em}
\begin{center}
Ural Federal University, Ekaterinburg, 620083, Russia
\end{center}

\vspace{1em}

\begin{abstract}
Phase diagrams of charged semi-hard-core bosons are studied in the mean field approximation. An increase in the parameter of local correlations is shown to lead to the transformation of the phase diagram of the system from the form characteristic of hard-core bosons to the limiting form with a parabolic dependence of the critical temperature of charge ordering on the boson concentration. The evolution between these limiting cases is dependent on the ratio between the model parameters and is accompanied by various effects, such as the change in the phase transition type, the appearance of new order-order transition, and the
appearance of new critical points.
\end{abstract}

\section{Introduction}
Recently the interest to the models of hard-core or semi-hard-core bosons~\cite{Dutta} increases because of the experimental observation of the competition of the charge ordering and the superconductivity in high-temperature superconductors~\cite{Demler} and the realization of cold atoms on an optical lattice~\cite{Islam}. A relatively long-living bound state of two bosons on an optical lattice was observed in~\cite{Winkler}. This fact makes topical the consideration of models with pair boson transfer, since it ceases to be the object of purely theoretical constructions. There are various versions~\cite{Heng, Sowinski, Zhou} of these models. In~\cite{Zhou}, in particular, the phase diagram of the model of semi-hard-core bosons was obtained taking into account two-boson transfer and local correlations. However, very little attention has been paid to the model without single-boson transfer~\cite{Panov}. In this work, we consider a system of charged bosons with a possible filling on a site $n = 0, 1, 2$ on a plane square lattice. In the framework of the pseudospin formalism \cite{Batista}, the Hamiltonian of the system can be written using the pseudospin operator $S = 1$ as follows~\cite{Moskvin, Panov2}:

\begin{equation}
	\hat{H} = \sum_i (\Delta S_{iz}^2 
	- \mu S_{iz}) + V \sum_{\left\langle ij\right\rangle} S_{iz} S_{jz} 
	- t_b \sum_{\left\langle ij\right\rangle} (S_{i+}^2S_{j-}^2+S_{i-}^2S_{j+}^2) . 
\label{eq:H00}
\end{equation}

Since the $z$ component of the pseudospin operator is
related to the operator of the number of bosons, $n_i = S_{iz} + 1$, the first term describes the effects of local
charge correlations on sites and the second term,
interstitial charge-charge correlations. In the third term, operators 
$B_{ix} = S_{i-}^2 + S_{i+}^2$ and $B_{iy} = i( S_{i-}^2 - S_{i+}^2 )$ are expressed in terms of creation (annihilation) of a pair of bosons on a site $S_{i+}^2$ $( S_{i-}^2 )$, where $S_{i\pm} = S_{ix} \pm iS_{iy}$. 
Taking into account the identity $B_{ix} B_{jx} + B_{iy} B_{jy} = 2( S_{i+}^2 S_{j-}^2 + S_{i-}^2 S_{j+}^2 )$, the third term is responsible for the transfers of boson pairs between the neighboring sites. The summation in the second and the third terms is performed over the nearest neighbors in a plane square
lattice. The last term proportional to chemical potential $\mu$ enables one to take into account the condition of
the constancy of the concentration. Further, we use
$x = n-1$, i.e., deviation of the concentration on the half filling instead $n$. 

The diagrams of the ground state of model~\eqref{eq:H00} in
the mean field approximation for various values of the
parameter of local correlations $\Delta$ and parameters of
charge-charge interaction $V$ and pair transfer t$_b$ were
constructed in~\cite{Panov}. 
It was shown that, at $\Delta \leq 0$, the diagram of the ground state of model~\eqref{eq:H00} is similar to the
case of hard-core bosons: at the $(V/t_b , x)$ plane to the
left of the Heisenberg point that for model~\eqref{eq:H00} is deter-
mined by relationship $2V/t_b = 1$, only superfluid (SF)
liquid is realized, and, to the right of this point, at
$x^2 < (2V/t_b - 1)/(2V/t_b + 1)$, the supersolid phase (SS)
is realized that transits to the charge ordering phase
(CO2) at the half-filling, $x = 0$. As $\Delta$ increases, the
regions of the SF and SS phase decrease and the
regions of CO phases increase; at $\Delta/t_b \geq 2$, there are
only three charge-ordered phases CO1, CO2, and
CO3 in the ground state. In~\cite{Panov2}, the mean field
approximation was used to study the concentration
dependences of the critical temperatures of the second-order transition for SF and CO phases. However, the first-order transitions are also realized in model~\eqref{eq:H00} at certain conditions. In addition, the well-known phase diagram of hard-core bosons~\cite{Robaszkiewicz} enables one to suggest the existence, in the phase diagram of semi-hard-core bosons, of various nontrivial situations, such as the change in the ordering type with an
increase in temperature, the phase separation (PS),
the existence of tricritical points, and the variety of
metastable states.

In this work, the evolution of phase diagrams of
model~\eqref{eq:H00} as a function of the parameter of local correlations $\Delta/t_b$ is studied in the mean field approximation for the most characteristic ratios $V/t_b$ . The paper
has the following structure. Section 2 gives a brief
description of the method of calculating the phase diagrams used in this work. Section 3 is devoted to an
analysis of the influence of local correlations on the
phase states of semi-hard-core bosons at various
values of the parameter of charge-charge correlations.
Section 4 contains brief conclusions.

\section{The mean field approximation}

Write the basic relations of the mean-field approximation obtained in~\cite{Panov2} which are necessary to construct the phase diagram of model~\eqref{eq:H00}. We use the Bogolyubov inequality to estimate the great potential of the system: $\Omega(H) \leq \Omega(H_0)+\langle H-H_0 \rangle_{H_0}$, where $H_0$ is the Hamiltonian of an ideal system. We introduce sublattices A and B which form the staggered order on a square lattice and write $H_0$ as
\begin{equation*}
	H_0=\sum_{c=1}^{N/2} H_c, \qquad H_c=H_A+H_B, 
\end{equation*}
\begin{equation}\label{H0}
H_{\alpha} = \Delta S_{z\alpha}^2-(h_z \pm h_z^a)S_{z\alpha}-(h_2 \pm h_2^a)B_{\alpha}.
\end{equation}
Here, $\alpha = A, B$ is the sublattice index, $h_z$, $h_z^a$, $h_2$ and $h_2^a$ are molecular fields that are variation parameters (vectors $\textbf{h}_2$, and $\textbf{h}_2^a$ have the $x$ and $y$ components). 
The partition function of an ideal system has the form
\begin{multline}
	Z_c = 4 \left(  1 
	+ e^{-\delta} \cosh \beta \sqrt{\left(h_z + h_z^a\right)^2 + \left(\textbf{h}_2 + \textbf{h}_2^a\right)^2} \right) \\
	\times \left( 1 
	+ e^{-\delta} \cosh \beta \sqrt{\left(h_z - h_z^a\right)^2 + \left(\textbf{h}_2 - \textbf{h}_2^a\right)^2} \right) .
\end{multline}
Here, $\beta = 1/k_B T$, where $k_B$ is the Boltzmann constant
(further we assume that $k_B = 1$), $T$ is temperature. This
enables us to write the expressions for $x$ and corresponding order parameters via molecular fields

\begin{eqnarray}
	x &=& \frac{1}{2\beta} \frac{\partial\ln Z_c}{\partial h_z}, 
	\label{eq:n} \\
	a &=& \frac{1}{2\beta} \frac{\partial\ln Z_c}{\partial h_z^a}, 
	\label{eq:a} \\
	\textbf{b} &=& \frac{1}{2\beta} \frac{\partial\ln Z_c}{\partial h_2}, 
	\label{eq:b} \\
	\textbf{b}_a &=& \frac{1}{2\beta} \frac{\partial\ln Z_c}{\partial h_2^a}. 
	\label{eq:ba}
\end{eqnarray}
and the estimation of the free energy per one site $f = \Omega/N + \mu x$:
\begin{multline}
\qquad
\qquad
	f = -\frac{1}{2\beta}\ln Z_c + 2V \left(n^2-a^2\right) - t_b \left(\textbf{b}^2-\textbf{b}_a^2\right) \\
	+ h_z n + h_z^a a + \textbf{h}_2 \textbf{b} + \textbf{h}_2^a \textbf{b}_a . 
	\label{eq:f}
\qquad
\qquad
\end{multline}

Minimizing the free energy, we obtain the equations
for the order parameters: 

\begin{equation}
	4 V a = h_z^a, \qquad 2t_b b = h_2, \qquad -2t_b b_a = h_2^a. 
\end{equation}

These equation should be solved numerically, taking into account Eqs.~(\ref{eq:n}--\ref{eq:ba}) at given $T$ and $x$. 
The obtained values of the order parameters and molecular fields allow us to calculate the free energy by Eq.~\eqref{eq:f}. The comparison of the free energy for various solutions enables us to build the phase diagram. The phase diagram regions corresponding to the phase separation were found using the Maxwell construction~\cite{Kapcia}: at given temperature $T$, boundary points $x_1$ and $x_2$ of the PS region are the solutions of the system of equations $\mu_1 (x_1 , T) = \mu (x_2 , T)$ and $\omega_1 (x_1 , T) = \omega_2 (x_2 , T)$, where $\mu_i$ is the chemical potential and $\omega_i$ is the specific grand potential of the $i$th phase.

\section{The peculiarities of the phase diagrams}

The system is characterized by the symmetry of the
phase diagrams with respect to $x = 0$; thus, it is sufficient to construct the phase diagrams for $0 \leq x \leq 0.5$.
Then, the phase diagram will be discussed taking
parameters $\Delta$, $V$, and $T $ in units of $t_b$.

The following types of the solutions of the system
of equation (9) are possible in the dependence on $T$, $x$,
and the relationships between the Hamiltonian parameters: high-temperature disordered phase (NO), in which all order parameters are zero; the superfluid (SF) phase with nonzero averages $\left\langle S_{A+}^2 \right\rangle = \left\langle S_{B+}^2 \right\rangle$, and order parameter $b \neq 0$; the charge ordering phase (CO) corresponding to the antiferromagnetic pseudospin ordering along axis z with order parameter $a \neq 0$; the supersolid (SS) phase, in which there are all order parameters $a \neq 0$, $b \neq 0$, $b_a \neq 0$; the phase separation (PS), i.e., state, in which the system is divided into macroscopic domains of the SF and CO phases.

In addition, we note that three types of the charge-ordered states can be in the ground state and at quite low temperatures~\cite{Panov}: CO1 and CO2 phases existing at $|x| < 1/2$, which differ in the character of filling of the sublattices, and CO3 for $0 < |x| < 1/2$. Let $p_{\alpha}(n)$ be the fraction of sites in sublattice $\alpha$ with the number of bosons $n$; we consider sets $p_{\alpha}(n) = \{ p_{\alpha}(0), p_{\alpha}(1), p_{\alpha}(2) \}$.
Then, at $0 \leqslant |x| \leqslant 1/2$, we have $p_A = \{ 0, 1, 0 \}$, $p_B = \{ 0, 1-2x, 2x \}$ in the CO1 phase and $p_A = \{ 1-2x, 2x,0 \} $, $p_B = \{ 0, 0, 1 \}$ in the CO2 phase. At $x = 1/2$,
phases CO1 and CO2 are transformed to the CO3 phase, for which $p_A = \{ 0, 2-2x , 2x-1 \}$, $p_B = \{ 0, 0, 1 \}$.

\begin{figure}
   \includegraphics[width=\textwidth]{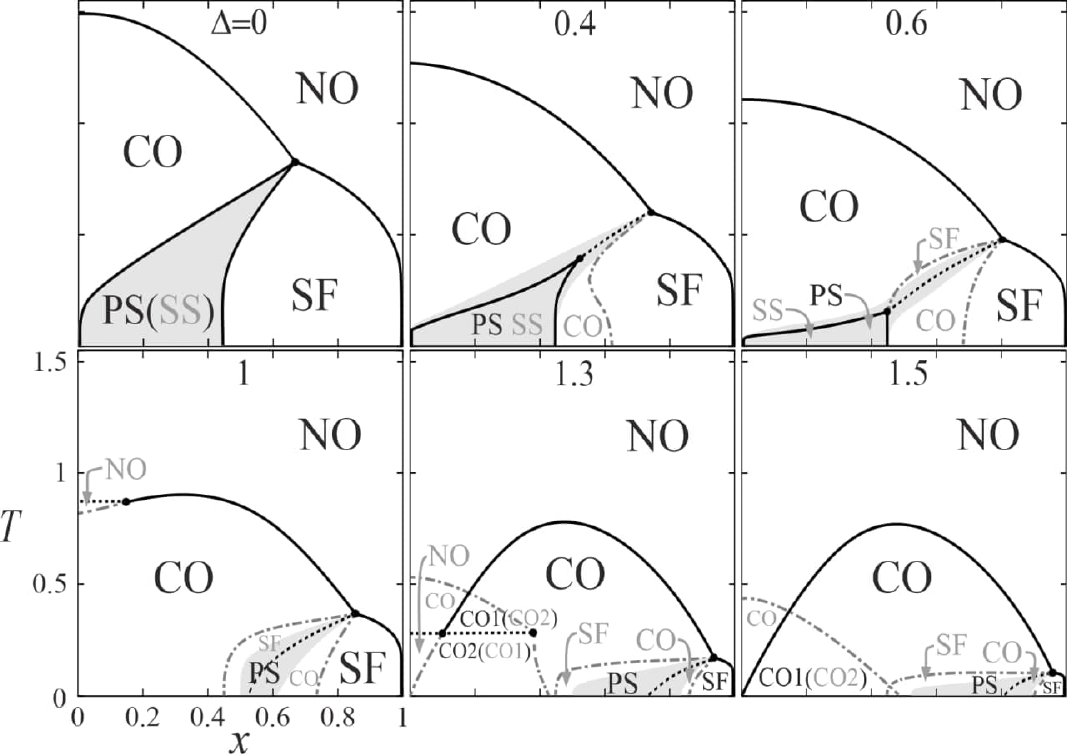} 
   \caption{Set of phase diagrams in the mean field approximation in the $x-T$ variables for a series of values of $\Delta$ and $V = 0.75$. The black solid (dotted) lines indicate the second (the first)-orders phase transitions; the grey color shows the phase separation region; the grey dash-dot line determines the phase stability boundaries.}
   \label{fig:fig1}
\end{figure}

Figure~\ref{fig:fig1} shows the set of the phase diagrams in the
$x-T$ variables for a series of values of $\Delta$ and $V = 0.75$.
The black solid and the black dotted lines indicate the
phase transitions of the second and first orders,
respectively; the grey dash-dot line shows the boundaries of stability of the coexisting phases; the grey color
shows the phase separation region. 

We can separate
two ranges of values of $\Delta$, for which the phase diagrams
are qualitatively different: $0 < \Delta < 0.75$ and $0.75 < \Delta < 1.5$. At $\Delta=0$, the phase diagram model~\eqref{eq:H00} is completely similar to the case of hard-core bosons~\cite{Robaszkiewicz}; in the first range of the local correlation parameters, mostly only quantitative changes are observed: as $\Delta$ increases, the maximum of the temperature of transition to CO phase $T_{CO}(x)$ is in point $x = 0$ and
decreases, and the maximum of $T_{SF}(x)$ corresponds to
the tricritical point (CO--NO--SF) and also decreases.
The stability region for the SF phase reduces, since the
positive values of $\Delta$ make states with $S_z = \pm 1$, the
exchange between which gives rise to the two-boson
transfer, unprofitable. The position of the PS region with respect to the CO and SF phases is not changed qualitatively, but the stability region of the SS-type solutions already does not coincide with the PS region, as in the case of hard-core bosons: the second tricritical point (CO--SS--SF) with lower $T$ and $x$
appears. It is important to note that, as in the case of
hard-core bosons, the free energy of the PS state is
always lower than the free energy of the SS phase. In
the second range $0.75 < \Delta < 1.5$, we observe already
qualitative distinctions in the shapes of the phase diagram from the case of hard-core bosons. Dependence $T_{CO}(x)$ becomes nonmonotonic, and the maximum begins to shift to $x = 0.5$. The region of existence
of the SF and PS phases is limited by values $0.5 < x < 1$,
and it reduces quickly. In this case, near $x = 0$, the second-order NO--CO transition is changed to the first-order transition, and the extension of the second-order transition line $T_{CO}(x)$ to the value at x = 0 determines the stability boundary of the metastable NO phase. The peculiarity of the first-order NO--CO phase transition near $x = 0$ is a relatively wide region of
stability of these phases. 
According to results of~\cite{Panov}, in model~\eqref{eq:H00}, at given
$V/t_b$ ratio, there is a critical value of $\Delta$, beginning from
which, in the ground state, phase CO2 is replaced by
the CO1 phase. The transformation of the phase diagram accompanying this replacement occurs nontrivially. Near $\Delta= 1.3$ , the order-order CO1--CO2 phase transition occurs at a temperature equal to that of the first-order NO--CO phase transition. The CO1--CO2
transition is also the first-order transition, which is confirmed by the existence of a finite jump of the charge parameter. The CO1--CO2 transition line ends in the critical point at $x_c < 0.5$. This transition temperature decreases with an increase in $\Delta$ and becomes zero at $\Delta=1.5$.

\begin{figure}
   \includegraphics[width=\textwidth]{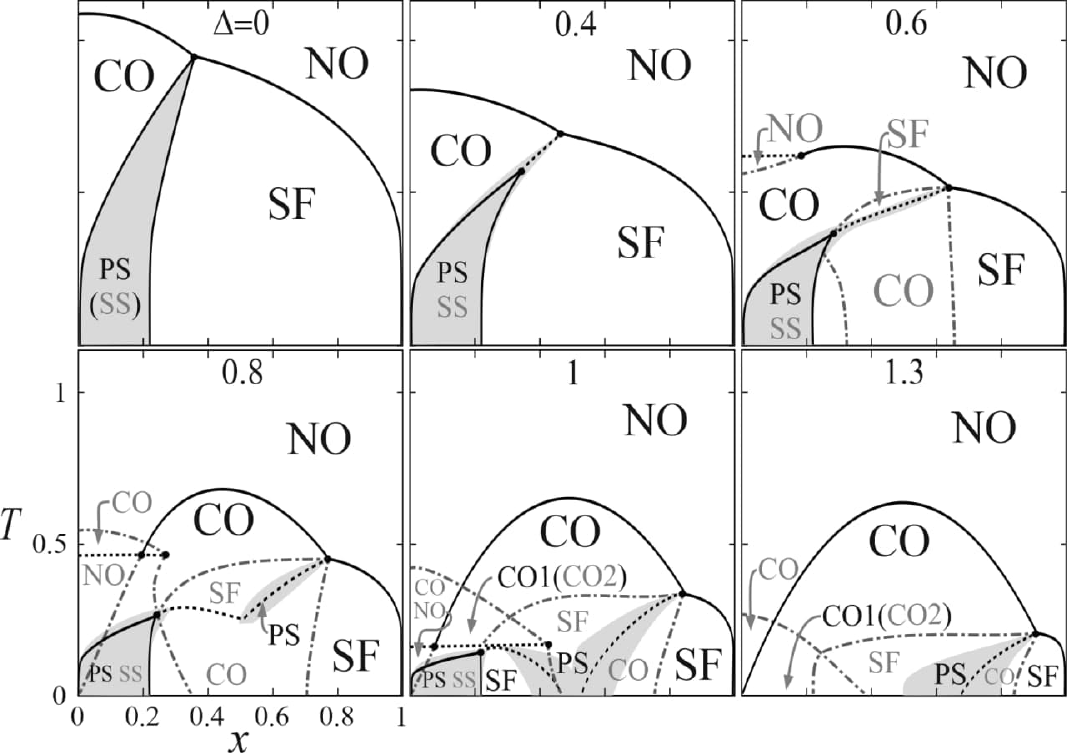} 
   \caption{Set of phase diagrams in the mean field approximation in the $x-T$ variables for a series of values of $\Delta$ and $V = 0.55$. The black solid (dotted) lines indicate the second (the first)-orders phase transitions; the grey color shows the phase separation region; the grey dash-dot line determines the phase stability boundaries.}
   \label{fig:fig2}
\end{figure}

\begin{figure}
   \includegraphics[width=\textwidth]{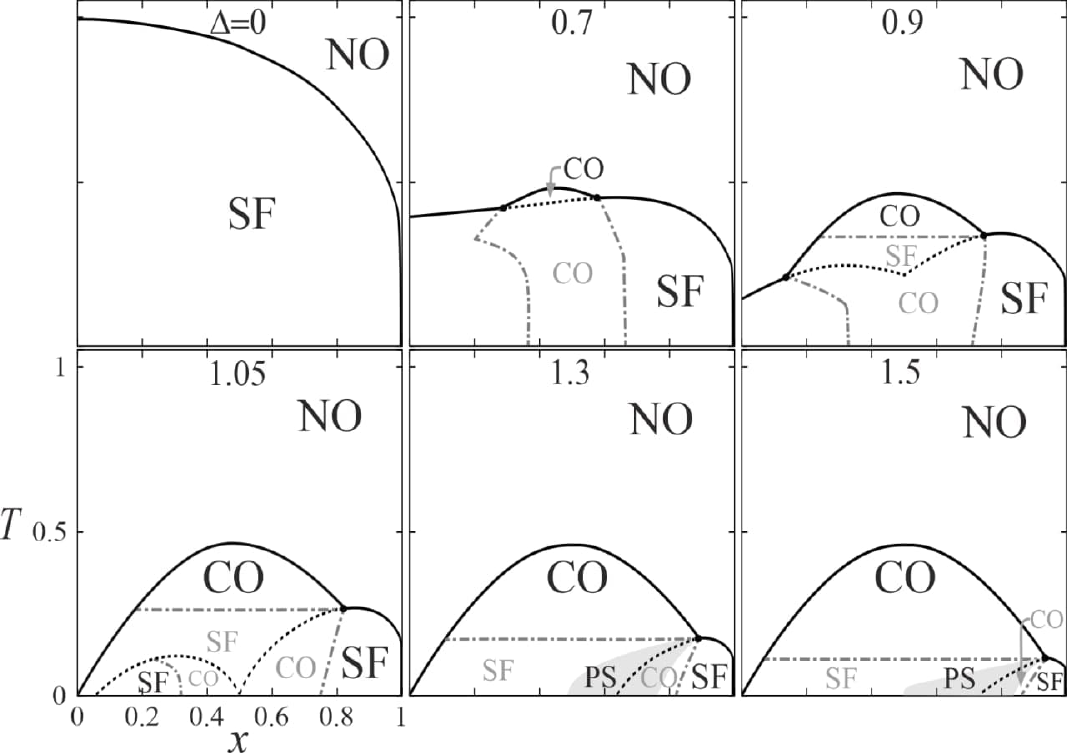} 
   \caption{Set of phase diagrams in the mean field approximation in the $x-T$ variables for a series of values of $\Delta$ and $V = 0.45$. The black solid (dotted) lines indicate the second (the first)-orders phase transitions; the grey color shows the phase separation region; the grey dash-dot line determines the phase stability boundaries.}
   \label{fig:fig3}
\end{figure}

Figures~\ref{fig:fig2} and~\ref{fig:fig3} demonstrate the evolution of the
phase diagrams with an increase in $\Delta$ at the values of $V$
close to the Heisenberg point $V = 0.5$. The case $V =
0.55$ shown in Fig.~\ref{fig:fig2} copies qualitatively the main
peculiarities of the phase diagrams at $V = 0.75$, but the
SF phase becomes more stable, particularly, at $\Delta < 1$.
There is interesting feature related to the phase separation: the PS-phase region unified at $\Delta= 0.6$, is separated into two unbound regions. The PS-phase region with $x > 0.5$ exists only at $T > 0$; however, a further increase in $\Delta$ stabilizes this type of the PS phase at low temperatures too. Figure~\ref{fig:fig3} shows the set of the phase diagrams for $V = 0.45$. At $\Delta = 0$, there are only two phases (NO and SF), but an increase in the parameter of local correlations $\Delta$ leads to the suppressing of the SF phase and the appearance of the CO phase in the phase diagram.
The CO phase appears first near $x = 0.5$ and gives the
way to the SF phase as temperature decreases. The
CO-SF transition is the first-order phase transition.
As $\Delta$ increases, the CO-phase region increases more
and more and replaces the SF phase; from some value
of $\Delta$, the phase separation region appears along the
phase equilibrium line of the CO and SF phases.

From Figs.~\ref{fig:fig1}--\ref{fig:fig3}, it is seen that at quite high $\Delta$ the
phase diagrams of model~\eqref{eq:H00} for all values of $V/t_b$ tend
to a unique universal type as phase CO is separated
from the NO phase by the second-order transition line
$T_{CO}(x)$ having the maximum at $x = 0.5$ and becoming
zero at $x = 0$ and $x = 1$. At $\Delta \rightarrow \infty$, the limiting dependence of the charge-ordering temperature is described
by the parabola~\cite{Panov2} $T_{CO}(x)/V = 4x(1 - x)$. The type 
of the limiting CO phase will correspond to CO1 at
 $|x| < 0.5$ and CO3 at $0.5 < |x| \leqslant 1$, since, in this case, the fraction of sites with $S_z^2 = 1$ is minimal. 

\section{Conclusions}

The phase diagrams for the system of charged semi-hard-core bosons were constructed using the numerical solution of the mean-field equations, and their evolution with a rise in the local correlation parameter $\Delta$ was studied. It is shown that, at $\Delta \leq 0$, the shape of the phase diagrams of model~\eqref{eq:H00} is completely similar to their shapes for the hard-core boson model; at $\Delta \rightarrow \infty$, the phase diagram of charged semi-hard-core bosons contains the only ordered CO phase, and the dependence of the charge ordering temperature on x is described by a parabola. The process of the transformation of the phase diagrams between these limiting cases is determined by the ratio of the parameter of interstitial charge-charge correlations to the integral of the two-boson transfer $V/t_b$. Below the Heisenberg point, at $V/t_b < 0.5$, the increase in $\Delta$ suppresses the SF phase and makes the CO phase more profitable at $x = 0.5$. Above the Heisenberg point, at $V/t_b > 0.5$, the evolution of the phase diagrams is more diversified: the regions of existence of all three initial phases CO, SF, and PS are changed as $\Delta$ increases. The evolution is accompanied by various nontrivial situations, such as the change in the phase transition type, the appearance of new order-order transitions, and the appearance of new critical points.

\bigskip

This work was supported by the Ministry of Education and Science of the Russian Federation (project no. FEUZ-2020-0054.

\end{document}